\documentclass[prl,aps,amssymb,amsfonts,amsmath,superscriptaddress,twocolumn]{revtex4}
\usepackage{graphicx, epstopdf} % Einbindung von Grafikdateien
\usepackage{amsfonts}%%% math fonts needed

\usepackage{graphicx,xcolor}
\usepackage{amsmath, bbm}
\usepackage{natbib}
\usepackage{braket}
\usepackage[utf8x]{inputenc}
\usepackage[ngerman, english]{babel}
\usepackage[T1]{fontenc}

\usepackage{hyperref}
\usepackage{subscript}

\newcommand{\mplaffil}{Max Planck Institute for the Science of Light (MPL), D-91058 Erlangen, Germany}
\newcommand{\fauaffil}{Department of Physics, Friedrich Alexander University of Erlangen-N\"urnberg, D-91058 Erlangen, Germany}
\newcommand{\leidenaffil}{Leiden Institute of Physics, PO Box 9504, 2300 RA Leiden, The Netherlands}

\begin{document} \title{Coherent Interaction of Light and Single Molecules \\ in a Dielectric Nanoguide }
\author{Sanli Faez}
\affiliation{\mplaffil}
\affiliation{\leidenaffil}
\author{Pierre T\"urschmann}
\affiliation{\mplaffil}
\author{Harald R. Haakh}
\affiliation{\mplaffil}
\author{Stephan G\"otzinger}
\affiliation{\fauaffil}
\affiliation{\mplaffil}
\author{Vahid Sandoghdar}
\affiliation{\mplaffil}
\affiliation{\fauaffil}

\begin{abstract} We present a new scheme for performing optical spectroscopy on single molecules. A glass capillary with a diameter of 600 nm filled with an organic crystal tightly guides the excitation light and provides a maximum spontaneous emission coupling factor ($\beta$) of 18\% for the dye molecules doped in the organic crystal. Combination of extinction, fluorescence excitation and resonance fluorescence spectroscopy with microscopy provides high-resolution spatio-spectral access to a very large number of single molecules in a linear geometry. We discuss strategies for exploring a range of quantum optical phenomena, including coherent cooperative interactions in a mesoscopic ensemble of molecules mediated by a single mode of propagating photons.\end{abstract} 
\pacs {42.50.-p, 42.50.Nn, 33.80.-b, 42.81.Qb}

\maketitle

The interaction of light and matter has played a prominent role in quantum physics and continues to kindle new areas of research and technology~\cite{Kimble:08}. The success of these emerging activities, however, will rely on the efficiency of the coupling between material and light particles. The most common route towards enhancing this interaction has been to trap and circulate photons in a high-finesse microcavity~\cite{Haroche-book:06}. In 2007 we demonstrated that it is even possible for propagating photons to reach a substantial coupling with individual quantum emitters directly~\cite{Gerhardt:07a}. The near-field arrangement of that experiment has been since extended to the far field, where tight focusing is shown to yield efficient interfacing of laser light with molecules, quantum dots, atoms or ions~\cite{Zumofen:08,Vamivakas:07,Wrigge:08,Tey:08,Pototschnig:11,Streed:12}. An intuitive explanation for such a strong cavity-free interaction is provided by the comparable values of $\sigma_0=3\lambda^2/2\pi$ as the resonant extinction cross section of a two-level atom and $(\lambda/2\rm sin\theta)^2$ as an approximate area for a diffraction-limited focal spot. Here $\lambda$ denotes the transition wavelength in the medium and $\theta$ stands for the half-angle that determines the numerical aperture (NA) of the focusing optics. Considering that a dielectric fiber with a subwavelength diameter can reach a mode area comparable with that of a tightly focused light beam \cite{Synder, Sandoghdar:97,Balykin:04}, it should also be possible to achieve an efficient interaction in this type of guided geometry. In what follows, we generally refer to a subwavelength waveguide as a `nanoguide'.

\begin{figure}[h]
\centering
\includegraphics[width=8.5cm]{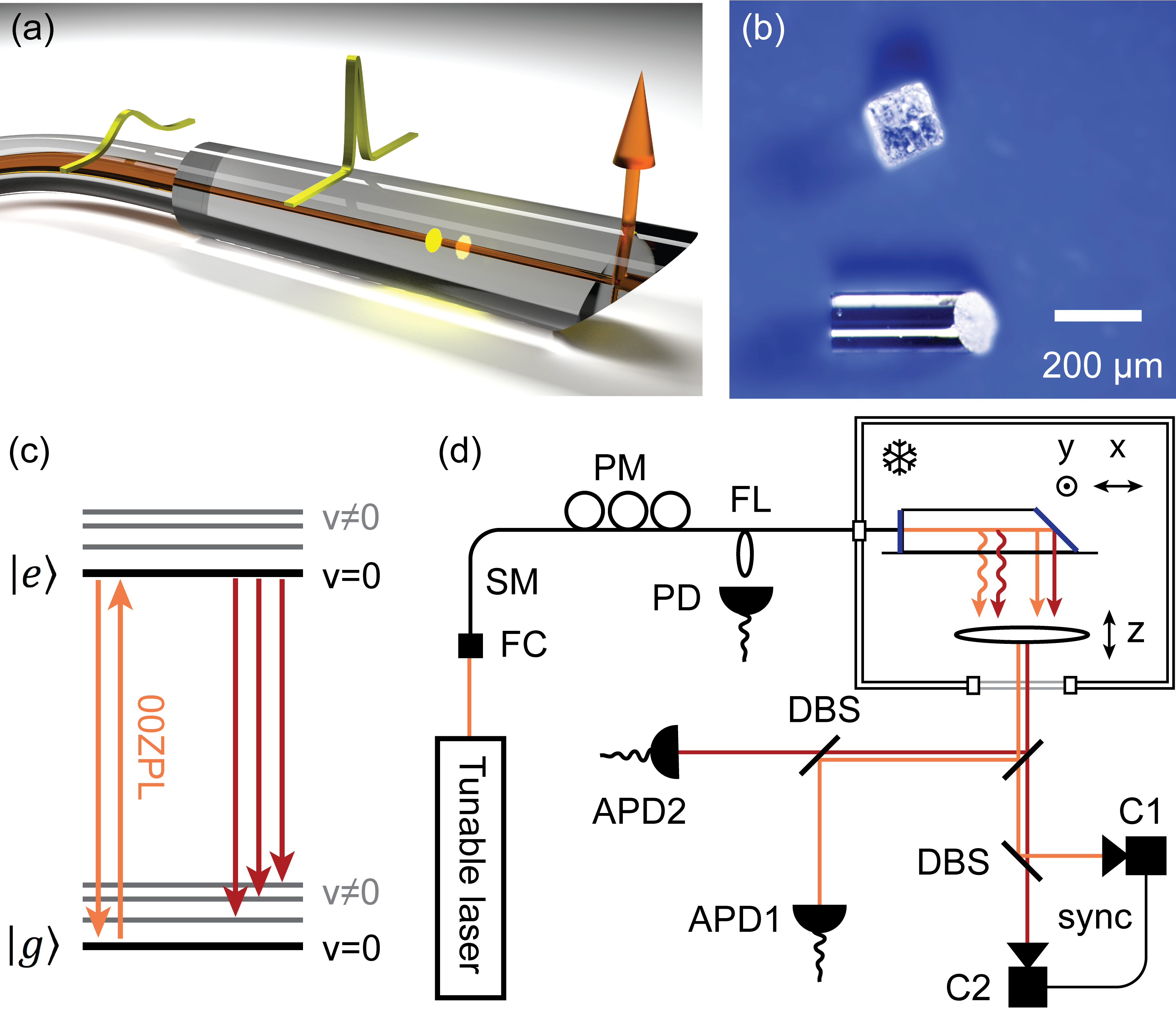}
\caption{ a) Schematics of a filled capillary butt-coupled to a single-mode optical fiber. b) A microscope image of an empty capillary next to a grain of table salt. c) Simplified Jablonski diagram of a DBT molecule. d) Sketch of the experimental arrangement. A CW Ti:Sapph laser is coupled to a single-mode fiber that enters a helium bath cryostat, ending with the nanocapillary. An aspheric lens provides optical access to the sample from a side window. Synchronized cameras C1 and C2 and avalanche photodiodes APD1 and APD2 are used for detection. PD: Photodiode; SM: single-mode fiber; DBS: dichromatic beam splitter; PM: fiber polarization management; FC: fiber coupler; FL: fiber leakage for intensity stabilization.}\label{setup}
\end{figure}

A comprehensive consideration beyond the discussion of the beam spot size and atomic cross section reveals that the key concept for maximizing the coupling efficiency is spatial mode matching~\cite{Zumofen:08}. In this picture, the effectiveness of a tightly focused beam is due to its large dipolar content, which is well adapted to the atomic emission pattern in free space~\cite{mojarad08}. Thus, an alternative strategy for optimizing the interaction could be to modify the spatial mode of the atomic radiation to match the profile of the incident light beam, for example, by employing parabolic~\cite{Sondermann:07} or planar~\cite{Lee:11} antennas. In this context, a nanoguide also acts as an antenna because it enhances the emission of an atom into its fundamental guiding mode~\cite{Chang:06,Friedler:09}. Interestingly, discussions of nanoguides have emerged somewhat in parallel to the tight focusing approach in different theoretical treatments~\cite{Shen:05,Chang:06,Friedler:09,hwang:11} and experimental studies of glass nanofibers~\cite{Balykin:04,vetsch_optical_2010,yalla_efficient_2012}, plasmonic nanowires~\cite{Akimov:07}, photonic crystal waveguides~\cite{lund-hansen_experimental_2008,Goban:14}, and even microwave transmission lines~\cite{Astafiev:10}. In this Letter we present a new system, where molecules are coupled to the optical mode of a nanoguide with a core made of an organic crystal surrounded by a thick glass cladding. This arrangement sets the ground for experiments on small ensembles of individual quantum emitters that are coupled via a photonic channel.

Figure~\ref{setup}a sketches the heart of our experimental setup. A fused silica capillary (refractive index $n_{2}=1.45$) with an inner diameter of 600~nm and an outer diameter of 160~$ \rm \mu$m was cut to a length of about 400~$ \rm \mu$m  with one cleaved end and a second polished facet at 45$^{\circ}$. Figure~\ref{setup}b shows a microscope image of an empty capillary next to a grain of table salt. The flat and angled ends of the nanocapillary were coated with 65~nm and 45~nm layers of aluminum, respectively. Once this central piece was fabricated, we heated it in the presence of a small amount of naphthalene ($n_{1}=1.59$) doped with dibenzoterrylene (DBT) molecules as quantum emitters at a nominal concentration of $10^{-7}$~M. After capillary forces filled the core, we cooled the sample and removed the residual naphthalene from the surrounding. Here, it is important that one obtains a homogeneous core to maintain a guiding mode and reduce scattering losses from grain boundaries. Since the vapor pressure of naphthalene is high at ambient conditions, the first few ten micrometers usually end up to be empty. Therefore, we filled this gap with ultraviolet curing glue (OG-142, Epotek).

DBT Molecules in the nanoguide can be spectrally isolated because as is common in solid-state media, the resonance frequencies of the individual molecules are distributed over an inhomogeneous spectrum \cite{Moerner:89}. The energy level scheme of DBT is illustrated in Fig.~\ref{setup}c. The transition of interest for this work is labeled 00ZPL, which involves the zero-phonon line (ZPL) of the transition between the ground vibrational levels (v=0) of the electronic ground ($\ket{g}$) and excited ($\ket{e}$) states. The 00ZPL of DBT is known to have a lifetime-limited linewidth of 30 MHz in naphthalene~\cite{Jelezko96}. However, the Frank-Condon and Debye-Waller factors of DBT reduce the resonant fluorescence contribution of 00ZPL by an overall branching ratio of $\alpha \approx 0.3$. To gain access to the narrow resonances of individual molecules, we used a tunable Ti:Sapphire ring laser with a linewidth of 100~kHz and performed the experiment at $T \lesssim 1.5 \rm K$ in a helium bath cryostat (see Fig.~\ref{setup}d).  

To couple the excitation laser light to the nanoguide, we fed the laser light into the cryostat through a single-mode optical fiber and butt-coupled it to the cleaved end of the capillary (see Fig.~\ref{setup}a). The aluminium coating at this end suppresses any scattering into the capillary cladding, which is caused by the mismatch between the mode areas of 14~$ \rm \mu m^2$ for the single-mode fiber and 0.3~$ \rm \mu m^2$ for the nanoguide. We covered the capillary by index-matched ultraviolet curing glue to achieve nearly distortion-free imaging of the nanoguide from the side and to fix the relative alignment of the capillary and the single-mode fiber on a cover glass. Nanopositioning systems mounted in the cryostat allowed us to position the nanoguide inside the asphere field of view of about 120~$\rm \mu m \times 120~\rm \mu$m and to adjust its focus (see Fig.~\ref{setup}d). In this fashion, we could record either the emission from the side of the nanoguide or transmission via the $45^\circ$ mirror at its end. 
  
\begin{figure}[h]
\centering
\includegraphics[width=8cm]{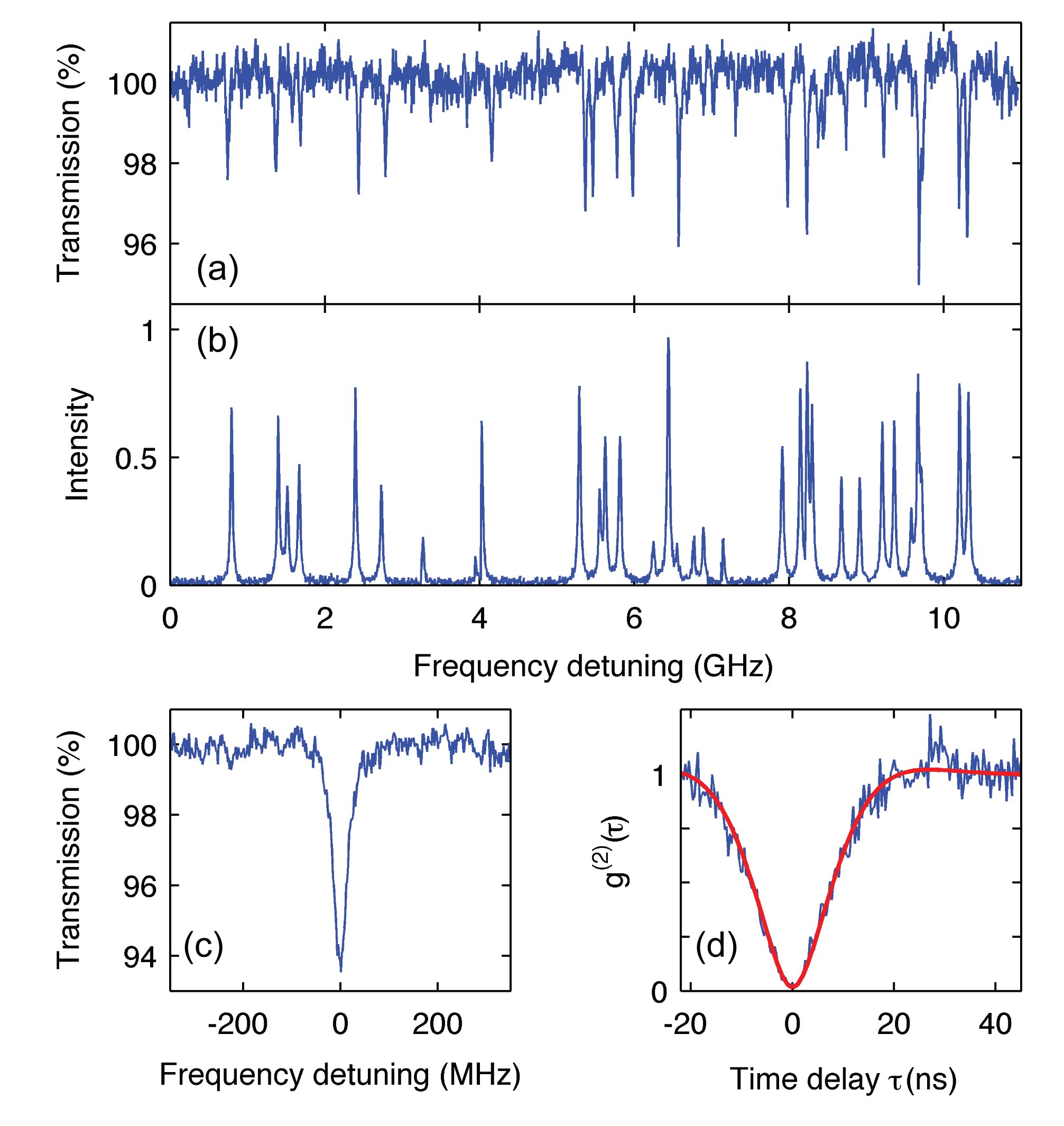}
\caption{a) Extinction spectrum recorded in transmission via the angled end of the nanoguide. We emphasize that these represent direct count rates without using lock-in detection. b) Stokes-shifted fluorescence spectrum recorded at the same time as (a). c) A zoom into an extinction dip of 6.5\% recorded from a single molecule. The excitation wavelength for the data in this figure was around 771 nm. d) Second-order autocorrelation function of the photons emitted on the resonance of (c). Here, we detected the Stokes-shifted fluorescence from the side. The recorded value of $g^{(2)}(0)=0.01$ was limited by the APD dark counts.} \label{peak-dip} 
\end{figure}

The polarization of the excitation light was adjusted by a paddle polarization controller. We sampled the intensity inside the single-mode fiber by inducing a small leakage (see Fig. \ref{setup}d) and used this signal to stabilize the excitation intensity against variations caused by thermal and mechanical perturbations. The light via the asphere was then sent to two synchronized detectors, which could be either sensitive cameras or avalanche photodiodes (APDs). In each case, a dichromatic beam splitter separated the excitation laser light from the Stokes-shifted fluorescence.  

Figure \ref{peak-dip}a displays an example of transmission spectra recorded through the angled output of the capillary. Extinction dips of a few percent are observed every time the resonance of a single molecule coincides with the scanned laser frequency. To verify this assignment, Fig.~\ref{peak-dip}b presents a spectrum recorded synchronously from the same port but using spectral filters to select the Stokes-shifted fluorescence. 

The extinguished power $\Delta P$ at each dip results from the interference of the incident light at power $P_0$ with the light that is resonantly scattered by an emitter into that mode. It follows that $\Delta P/P_0=2\beta-\beta^2$, where $\beta$ stands for the fraction of the dipolar emission into the nanoguide mode compared to its overall radiation~\cite{Haakh:14, EPAPS-Tuerschmann}. Accounting for the internal branching ratio $\alpha$ of the molecule, one arrives then at $\Delta P/P_0=2(\alpha\beta)-(\alpha\beta)^2$. In Fig.~\ref{peak-dip}c we show an example of an extinction dip reaching 6.5\%, or equivalently $\beta \approx 11\%$. To verify that this signal stems from a single molecule, we performed Hanbury-Brown and Twiss coincidence measurements on this resonance. The outcome presented in Fig.~\ref{peak-dip}d confirms antibunched emission with a lifetime of 5.3 ns, corresponding to the measured linewidth of $\gamma=30$~MHz.

As shown in Fig.~\ref{peak-dip}a, the recorded extinction dips display a certain degree of variation and, in fact, we have observed dips as large as 9\%. We attribute this distribution to the radial position of the molecules and the orientation of their dipole moments. Considering that the volume of a cylindrical shell at a distance $r$ from the center grows linearly with $r$, a smaller fraction of the molecules experience the highest optical coupling. 

\begin{figure}[b!]
\centering
\includegraphics[width=6.5cm]{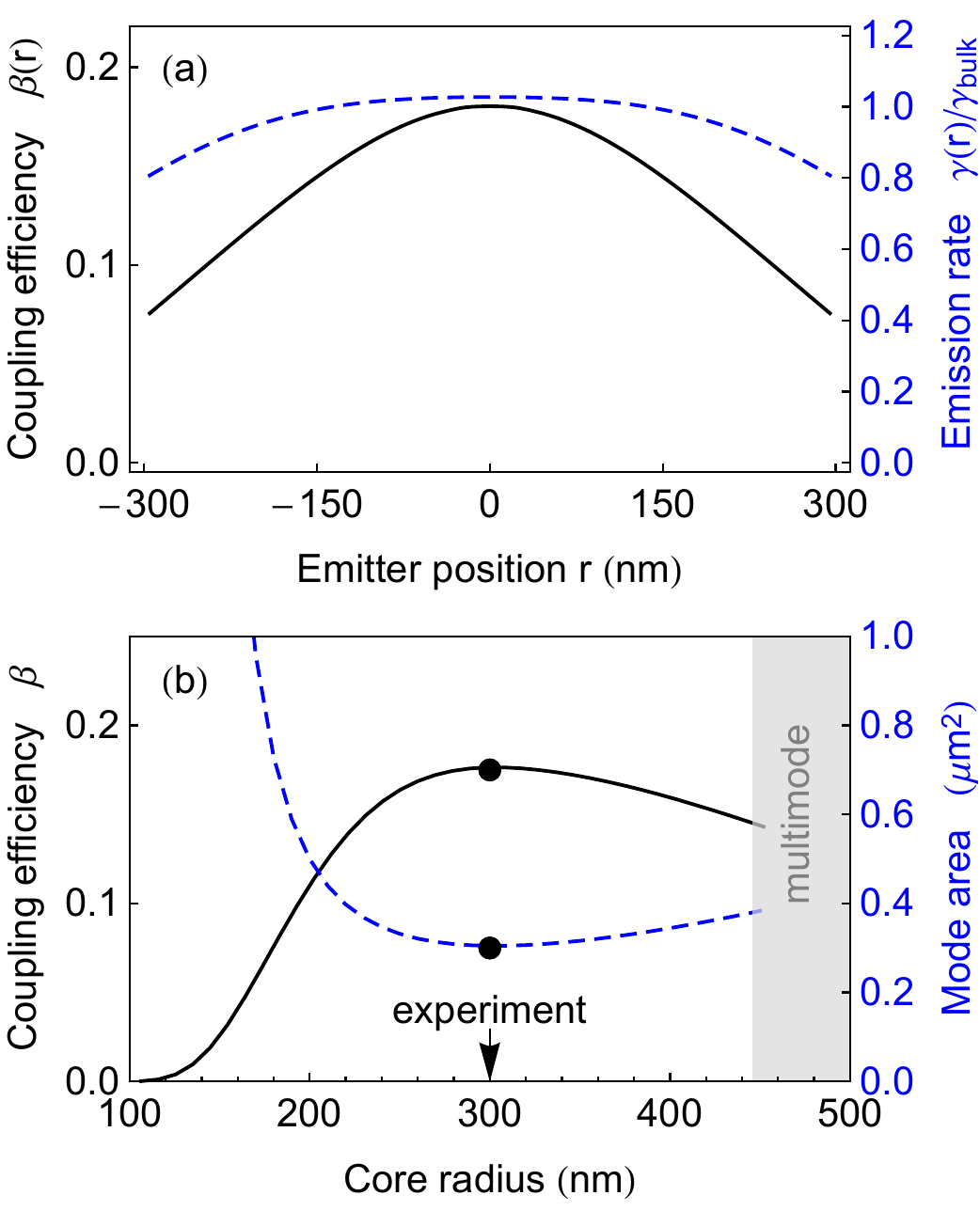}
\caption{a) Nanoguide-emitter coupling efficiency $\beta$ (black solid curve, left axis) and the total decay rate (blue dashed curve, right axis) as a function of the emitter position inside a nanocapillary core of 600 nm. b) $\beta$ as a function of the core radius (black solid curve, left axis). The blue dashed trace plots the effective mode area (right axis). The dot indicates the optimal core radius $r =300~{\rm nm}$, which was used in the experiment. The nanocapillary core was assumed to have a refractive index of $n_1 = 1.59$ surrounded by a large cladding with $n_2=1.45$. The shaded region in (b) marks the regime, where the waveguide becomes multimode.} \label{beta}
\end{figure}

The black curve in Fig.~\ref{beta}a plots the calculated values of $\beta$ for radially oriented dipole moments as a function of $r$. The expected range of $\beta$=7-18\% translates to extinction dips of  4-11\% in our setup when considering $\alpha$. Finally, we emphasize that the inner radius of the capillary in our experiment was chosen to yield the highest coupling for the refractive indices of a naphthalene core and a glass cladding. The results of calculations reported in Fig.~\ref{beta}b show the variation of $\beta$ and the mode area as a function of the capillary inner radius. 

To present a thorough consideration of the radiative properties of a molecule coupled to the nanoguide, we also calculated the cavity quantum electrodynamic effect of the boundaries. The dashed blue curve in Fig.~\ref{beta}a indicates that the spontaneous emission rate of a radially-oriented dipolar emitter changes by less than 20\% across the nanoguide diameter. This level of variation is negligible in our work because the linewidths of dye molecules in organic hosts commonly show a larger distribution~\cite{SMbook}. In fact, we suspect that the crystallization process in the highly confined space of a nanocapillary might lead to additional mechanical stress and spectral inhomogeneity. We present a distribution of the measured linewidths in Fig. S2 of the online supplementary material~\cite{EPAPS-Tuerschmann}, but more detailed investigations of the homogeneous and inhomogeneous spectra in this novel geometry requires further studies.

\begin{figure}[t]
\centering
\includegraphics[width=8cm]{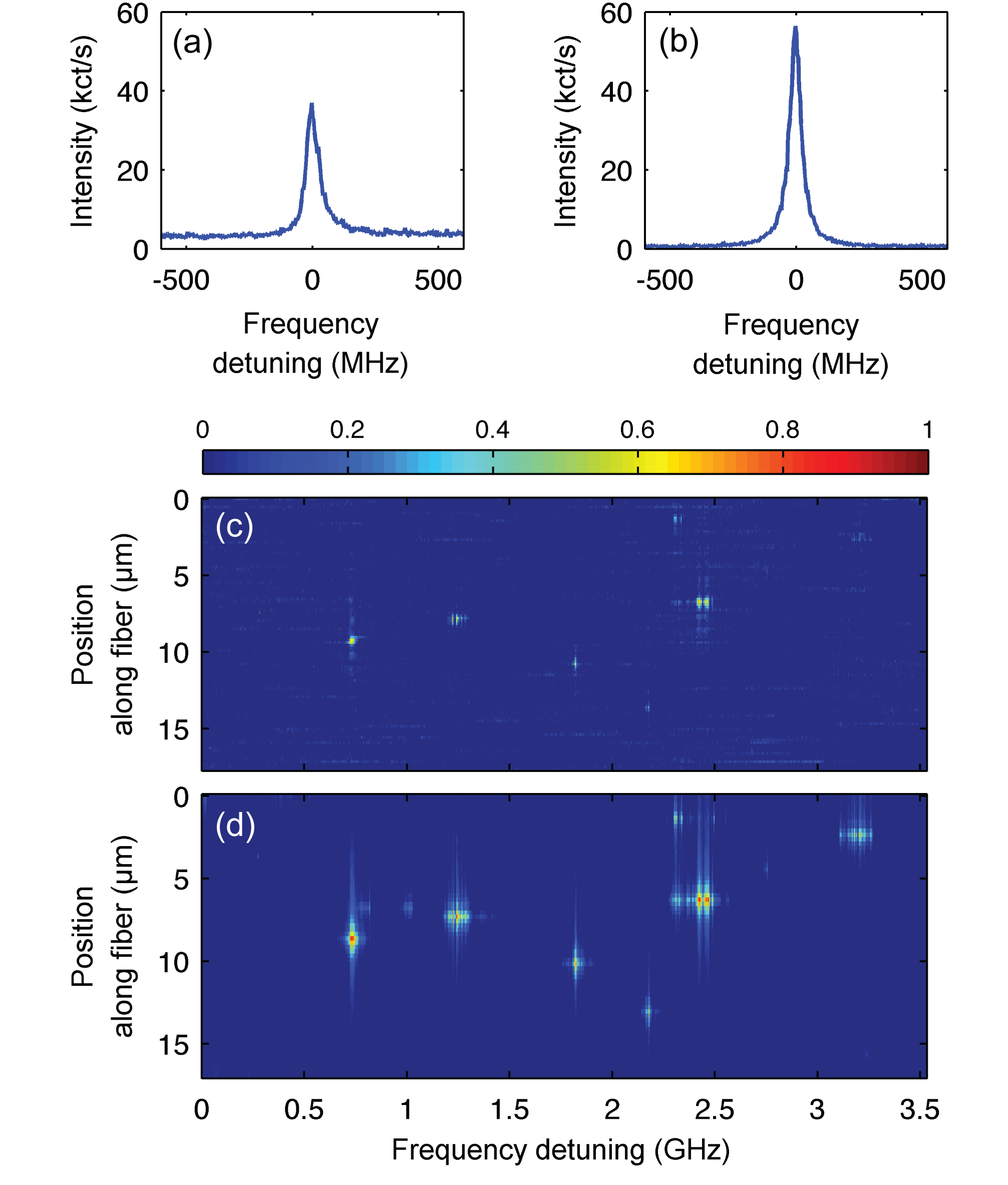}
\caption{a) A resonance fluorescence spectrum of a single DBT molecule recorded from the side. b) The Stokes-shifted fluorescence spectrum of the same molecule. c) A section of the resonance fluorescence excitation spatio-spectral map of the sample. d) The same for the Stokes-shifted signal recorded simultaneously. The data in this figure were recorded at an excitation wavelength around 757 nm. The linear color scales represent the signal strength in arbitrary units.} \label{red_res} 
\end{figure}

Aside from achieving a high efficiency, our experimental arrangement provides access to the resonance fluorescence of molecules on a low background. Figure~\ref{red_res}a displays an example of a single-molecule excitation spectrum obtained from the side of the nanoguide through a bandpass filter that transmits the excitation wavelength. This detection scheme offers a novel configuration for right-angle spectroscopy on molecules in a matrix. Although such measurements have been standard in gas-phase atomic and molecular physics~\cite{Walther:05}, they are very challenging in the solid state due to a strong background scattering. Figure~\ref{red_res}b presents the spectrum of the same molecule as in Fig.~\ref{red_res}a but detected through its Stokes-shifted fluorescence from the side. Monitoring the polarization of the detected fluorescence, we verified that the dipole moments of the molecules were generally aligned at an angle of about $80^\circ$ relative to the nanoguide axis. However, rotation of the incident polarization within the nanoguide indicated that the radial dipole component did not have a fixed direction. 

The side port of the nanoguide also brings about the important feature that all the molecules across the thin capillary core can be imaged at the same time since they lie within the depth of focus of the aspheric lens. Figures~\ref{red_res}c and d show spatio-spectral maps of the resonance and Stokes-shifted fluorescence along more than 15~$\mu$m of the nanoguide and within a spectral range of about 3 GHz. Considering that one can localize the center of the dipole moment of each molecule with \r angstrom precision through the analysis of its image point-spread function~\cite{Weisenburger:14}, the data in Fig.~\ref{red_res} illustrate the ability to address each molecule through high-resolution spectral and spatial information. 

A noteworthy aspect of the molecules in the nanoguide is that they are all well within a coherence length $l_c$ of each other, where $l_c=\tau c/n_1 \approx 1$~m. As a result, far-field single-mode interaction of two or more molecules can lead to a coupling strength of $\gamma$ for $\beta=1$ ~\cite{Gonzalez-Tudela:11} in analogy with effects in near-field dipole-dipole coupling~\cite{Hettich:02}. First steps towards the realization of these ideas have been taken using two molecules via free space~\cite{Rezus:12} and two superconducting qubits along a waveguide~\cite{vanLoo:13}. The system presented here offers great promise for coupling a small ensemble of individual molecules that have been carefully selected. 

\begin{figure}[h]
\centering
\includegraphics[width=8cm]{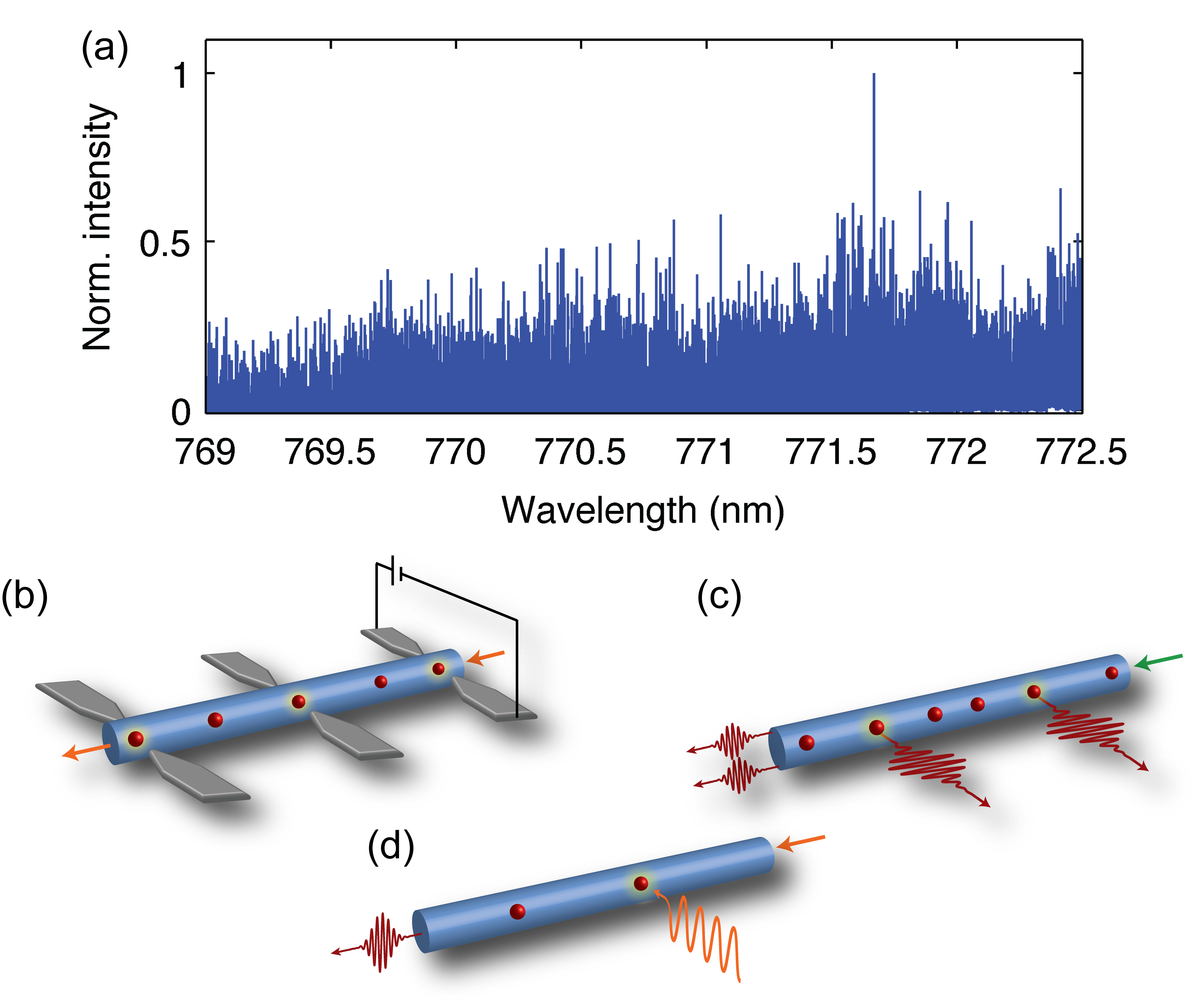}
\caption{a) Stokes-shifted excitation spectrum recorded in transmission. The inhomogeneous distribution of DBT lines in our system is considerably larger than that observed in sublimated naphthalene crystals. Furthermore, the current site in the region of 770 nm has not been reported previously. \mbox{ b) Schematics } of micro/nano electrodes for controlling the resonance frequencies of the molecules along the nanoguide. \mbox{c) Arrangement} for generating several indistinguishable photons which would be detected either from the side port or the end. Here, one would pump the medium via v=1 vibrational level of the electronic excited state. \mbox{d)} This cartoon illustrates resonant excitation and detection of the 00ZPL exploiting a right-angle geometry.} \label{outlook} 
\end{figure}

Figure~\ref{outlook}a plots an excitation spectrum of about 5000 molecules recorded in transmission, displaying a large spectral inhomogeneity. A powerful feature of the nanoguide geometry is that contrary to high-NA optics with small depth of focus, a guided photon can couple to a large number of resonant molecules along its length. Application of local electric fields via micro- and nano-electrodes (see Fig.~\ref{outlook}b) combined with side-port spectro-spatial mapping will allow one to tune a subset of these molecules into resonance~\cite{Rezus:12}. Similarly, one can remove accidental resonance of several molecules along the nanoguide in this fashion. 

The possibility of exciting two or more molecules at the same time would also provide an opportunity to generate several indistinguishable photons either in transmission or via the side port (see Fig.~\ref{outlook}c). Figure~\ref{outlook}d sketches another study, where one can take advantage of the side port to control the population of the excited state of a single molecule resonantly via 00ZPL~\cite{Gerhardt:09} without the need for excitation to higher vibrational states, which are more than thousand times broader~\cite{Hwang:09,Rezus:12}. This feature can be employed to generate narrow-band single photons or gate a single-molecule transistor~\cite{Hwang:09} using only very few photons. 

In conclusion, the experimental arrangement and ideas presented in this Letter provide a versatile and efficient platform for realizing long-distance photonic interactions among many quantum emitters along a bus. Indeed, an increasing number of theoretical reports have recently pointed out intriguing phenomena in related systems, ranging from multiphoton nonlinear processes~\cite{Shen:07} to spin-charge separation~\cite{Greentree:06}. Many of these experiments would greatly benefit from larger $\beta$ factors, which can be pursued in different architectures such as photonic crystals~\cite{lund-hansen_experimental_2008,Goban:14} or slot waveguides~\cite{Quan:09}. In some other studies, moderate $\beta$ factors could suffice to explore novel polaritonic phenomena. 

\bigskip

We acknowledge financial support from the European Union (ERC Advanced Grant SINGLEION) and the Max Planck Society. We also thank the Fiber Technology Development and Service Unit at MPL for providing the glass capillaries and Patricia Schrehardt for cutting and polishing them. S.F. and P.T. contributed equally to this work.

\bigskip


\begin{thebibliography}{99}
\expandafter\ifx\csname natexlab\endcsname\relax\def\natexlab#1{#1}\fi
\expandafter\ifx\csname bibnamefont\endcsname\relax
  \def\bibnamefont#1{#1}\fi
\expandafter\ifx\csname bibfnamefont\endcsname\relax
  \def\bibfnamefont#1{#1}\fi
\expandafter\ifx\csname citenamefont\endcsname\relax
  \def\citenamefont#1{#1}\fi
\expandafter\ifx\csname url\endcsname\relax
  \def\url#1{\texttt{#1}}\fi
\expandafter\ifx\csname urlprefix\endcsname\relax\def\urlprefix{URL }\fi
\providecommand{\bibinfo}[2]{#2}
\providecommand{\eprint}[2][]{\url{#2}}

\bibitem[{\citenamefont{Kimble}(2012)}]{Kimble:08}
\bibinfo{author}{\bibfnamefont{H.~J.} \bibnamefont{Kimble}},
  \bibinfo{journal}{Nature} \textbf{\bibinfo{volume}{453}},
  \bibinfo{pages}{1023} (\bibinfo{year}{2012}).

\bibitem[{\citenamefont{Raimond and Haroche}(2006)}]{Haroche-book:06}
\bibinfo{author}{\bibfnamefont{J.-M.} \bibnamefont{Raimond}} \bibnamefont{and}
  \bibinfo{author}{\bibfnamefont{S.}~\bibnamefont{Haroche}},
  \emph{\bibinfo{title}{Exploring the Quantum}} (\bibinfo{publisher}{Oxford
  University Press}, \bibinfo{address}{Oxford, UK}, \bibinfo{year}{2006}).

\bibitem[{\citenamefont{Gerhardt et~al.}(2007)\citenamefont{Gerhardt, Wrigge,
  Bushev, Zumofen, Agio, Pfab, and Sandoghdar}}]{Gerhardt:07a}
\bibinfo{author}{\bibfnamefont{I.}~\bibnamefont{Gerhardt}},
  \bibinfo{author}{\bibfnamefont{G.}~\bibnamefont{Wrigge}},
  \bibinfo{author}{\bibfnamefont{P.}~\bibnamefont{Bushev}},
  \bibinfo{author}{\bibfnamefont{G.}~\bibnamefont{Zumofen}},
  \bibinfo{author}{\bibfnamefont{M.}~\bibnamefont{Agio}},
  \bibinfo{author}{\bibfnamefont{R.}~\bibnamefont{Pfab}}, \bibnamefont{and}
  \bibinfo{author}{\bibfnamefont{V.}~\bibnamefont{Sandoghdar}},
  \bibinfo{journal}{Phys. Rev. Lett.} \textbf{\bibinfo{volume}{98}},
  \bibinfo{pages}{033601} (\bibinfo{year}{2007}).

\bibitem[{\citenamefont{Zumofen et~al.}(2008)\citenamefont{Zumofen, Mojarad,
  Sandoghdar, and Agio}}]{Zumofen:08}
\bibinfo{author}{\bibfnamefont{G.}~\bibnamefont{Zumofen}},
  \bibinfo{author}{\bibfnamefont{N.~M.} \bibnamefont{Mojarad}},
  \bibinfo{author}{\bibfnamefont{V.}~\bibnamefont{Sandoghdar}},
  \bibnamefont{and} \bibinfo{author}{\bibfnamefont{M.}~\bibnamefont{Agio}},
  \bibinfo{journal}{Phys. Rev. Lett.} \textbf{\bibinfo{volume}{101}},
  \bibinfo{pages}{180404} (\bibinfo{year}{2008}).

\bibitem[{\citenamefont{Vamivakas et~al.}(2007)\citenamefont{Vamivakas,
  Atat\"ure, Dreiser, Yilmaz, Badolato, Swan, Goldberg, Imamoglu, and
  \"Unl\"u}}]{Vamivakas:07}
\bibinfo{author}{\bibfnamefont{A.~N.} \bibnamefont{Vamivakas}},
  \bibinfo{author}{\bibfnamefont{M.}~\bibnamefont{Atat\"ure}},
  \bibinfo{author}{\bibfnamefont{J.}~\bibnamefont{Dreiser}},
  \bibinfo{author}{\bibfnamefont{S.~T.} \bibnamefont{Yilmaz}},
  \bibinfo{author}{\bibfnamefont{A.}~\bibnamefont{Badolato}},
  \bibinfo{author}{\bibfnamefont{A.~K.} \bibnamefont{Swan}},
  \bibinfo{author}{\bibfnamefont{B.~B.} \bibnamefont{Goldberg}},
  \bibinfo{author}{\bibfnamefont{A.}~\bibnamefont{Imamoglu}}, \bibnamefont{and}
  \bibinfo{author}{\bibfnamefont{M.~S.} \bibnamefont{\"Unl\"u}},
  \bibinfo{journal}{Nano Letters} \textbf{\bibinfo{volume}{7}},
  \bibinfo{pages}{2892} (\bibinfo{year}{2007}).

\bibitem[{\citenamefont{Wrigge et~al.}(2008)\citenamefont{Wrigge, Gerhardt,
  Hwang, Zumofen, and Sandoghdar}}]{Wrigge:08}
\bibinfo{author}{\bibfnamefont{G.}~\bibnamefont{Wrigge}},
  \bibinfo{author}{\bibfnamefont{I.}~\bibnamefont{Gerhardt}},
  \bibinfo{author}{\bibfnamefont{J.}~\bibnamefont{Hwang}},
  \bibinfo{author}{\bibfnamefont{G.}~\bibnamefont{Zumofen}}, \bibnamefont{and}
  \bibinfo{author}{\bibfnamefont{V.}~\bibnamefont{Sandoghdar}},
  \bibinfo{journal}{Nature Phys.} \textbf{\bibinfo{volume}{4}},
  \bibinfo{pages}{60} (\bibinfo{year}{2008}).

\bibitem[{\citenamefont{Tey et~al.}(2008)\citenamefont{Tey, Chen, Aljunid,
  Chng, Huber, Maslennikov, and Kurtsiefer}}]{Tey:08}
\bibinfo{author}{\bibfnamefont{M.~K.} \bibnamefont{Tey}},
  \bibinfo{author}{\bibfnamefont{Z.}~\bibnamefont{Chen}},
  \bibinfo{author}{\bibfnamefont{S.~A.} \bibnamefont{Aljunid}},
  \bibinfo{author}{\bibfnamefont{B.}~\bibnamefont{Chng}},
  \bibinfo{author}{\bibfnamefont{F.}~\bibnamefont{Huber}},
  \bibinfo{author}{\bibfnamefont{G.}~\bibnamefont{Maslennikov}},
  \bibnamefont{and}
  \bibinfo{author}{\bibfnamefont{C.}~\bibnamefont{Kurtsiefer}},
  \bibinfo{journal}{Nature Phys.} \textbf{\bibinfo{volume}{924}},
  \bibinfo{pages}{4} (\bibinfo{year}{2008}).

\bibitem[{\citenamefont{Pototschnig et~al.}(2011)\citenamefont{Pototschnig,
  Chassagneux, Hwang, Zumofen, Renn, and Sandoghdar}}]{Pototschnig:11}
\bibinfo{author}{\bibfnamefont{M.}~\bibnamefont{Pototschnig}},
  \bibinfo{author}{\bibfnamefont{Y.}~\bibnamefont{Chassagneux}},
  \bibinfo{author}{\bibfnamefont{J.}~\bibnamefont{Hwang}},
  \bibinfo{author}{\bibfnamefont{G.}~\bibnamefont{Zumofen}},
  \bibinfo{author}{\bibfnamefont{A.}~\bibnamefont{Renn}}, \bibnamefont{and}
  \bibinfo{author}{\bibfnamefont{V.}~\bibnamefont{Sandoghdar}},
  \bibinfo{journal}{Phys. Rev. Lett.} \textbf{\bibinfo{volume}{107}},
  \bibinfo{pages}{063001} (\bibinfo{year}{2011}).

\bibitem[{\citenamefont{Streed et~al.}(2012)\citenamefont{Streed, Jechow,
  Norton, and Kielpinski}}]{Streed:12}
\bibinfo{author}{\bibfnamefont{E.~W.} \bibnamefont{Streed}},
  \bibinfo{author}{\bibfnamefont{A.}~\bibnamefont{Jechow}},
  \bibinfo{author}{\bibfnamefont{B.~G.} \bibnamefont{Norton}},
  \bibnamefont{and}
  \bibinfo{author}{\bibfnamefont{D.}~\bibnamefont{Kielpinski}},
  \bibinfo{journal}{Nat. Commun.} \textbf{\bibinfo{volume}{3}},
  \bibinfo{pages}{933} (\bibinfo{year}{2012}).

\bibitem[{\citenamefont{Synder and Love}(1983)}]{Synder}
\bibinfo{author}{\bibfnamefont{A.~W.} \bibnamefont{Synder}} \bibnamefont{and}
  \bibinfo{author}{\bibfnamefont{J.~D.} \bibnamefont{Love}},
  \emph{\bibinfo{title}{Optical Waveguide Theory}} (\bibinfo{publisher}{Chapman
  \& Hall}, \bibinfo{address}{London, UK}, \bibinfo{year}{1983}).

\bibitem[{\citenamefont{Sandoghdar et~al.}(1997)\citenamefont{Sandoghdar,
  Wegscheider, Krausch, and Mlynek}}]{Sandoghdar:97}
\bibinfo{author}{\bibfnamefont{V.}~\bibnamefont{Sandoghdar}},
  \bibinfo{author}{\bibfnamefont{S.}~\bibnamefont{Wegscheider}},
  \bibinfo{author}{\bibfnamefont{G.}~\bibnamefont{Krausch}}, \bibnamefont{and}
  \bibinfo{author}{\bibfnamefont{J.}~\bibnamefont{Mlynek}},
  \bibinfo{journal}{J. Appl. Phys.} \textbf{\bibinfo{volume}{81}},
  \bibinfo{pages}{2499} (\bibinfo{year}{1997}).

\bibitem[{\citenamefont{Balykin et~al.}(2004)\citenamefont{Balykin, Hakuta,
  Kien, Liang, and Morinaga}}]{Balykin:04}
\bibinfo{author}{\bibfnamefont{V.~I.} \bibnamefont{Balykin}},
  \bibinfo{author}{\bibfnamefont{K.}~\bibnamefont{Hakuta}},
  \bibinfo{author}{\bibfnamefont{F.~L.} \bibnamefont{Kien}},
  \bibinfo{author}{\bibfnamefont{J.~Q.} \bibnamefont{Liang}}, \bibnamefont{and}
  \bibinfo{author}{\bibfnamefont{M.}~\bibnamefont{Morinaga}},
  \bibinfo{journal}{Phys. Rev. A.} \textbf{\bibinfo{volume}{70}},
  \bibinfo{pages}{011401(R)} (\bibinfo{year}{2004}).

\bibitem[{\citenamefont{Mojarad et~al.}(2008)\citenamefont{Mojarad, Sandoghdar,
  and Agio}}]{mojarad08}
\bibinfo{author}{\bibfnamefont{N.~M.} \bibnamefont{Mojarad}},
  \bibinfo{author}{\bibfnamefont{V.}~\bibnamefont{Sandoghdar}},
  \bibnamefont{and} \bibinfo{author}{\bibfnamefont{M.}~\bibnamefont{Agio}},
  \bibinfo{journal}{J. Opt. Soc. Am. B} \textbf{\bibinfo{volume}{25}},
  \bibinfo{pages}{651} (\bibinfo{year}{2008}).

\bibitem[{\citenamefont{Sondermann et~al.}(2007)\citenamefont{Sondermann,
  Maiwald, Konermann, Lindlein, Peschel, and Leuchs}}]{Sondermann:07}
\bibinfo{author}{\bibfnamefont{M.}~\bibnamefont{Sondermann}},
  \bibinfo{author}{\bibfnamefont{R.}~\bibnamefont{Maiwald}},
  \bibinfo{author}{\bibfnamefont{H.}~\bibnamefont{Konermann}},
  \bibinfo{author}{\bibfnamefont{N.}~\bibnamefont{Lindlein}},
  \bibinfo{author}{\bibfnamefont{U.}~\bibnamefont{Peschel}}, \bibnamefont{and}
  \bibinfo{author}{\bibfnamefont{G.}~\bibnamefont{Leuchs}},
  \bibinfo{journal}{Appl. Phys. B} \textbf{\bibinfo{volume}{89}},
  \bibinfo{pages}{489} (\bibinfo{year}{2007}).

\bibitem[{\citenamefont{Lee et~al.}(2011)\citenamefont{Lee, Chen, Eghlidi,
  Kukura, Lettow, Renn, Sandoghdar, and G{\"o}tzinger}}]{Lee:11}
\bibinfo{author}{\bibfnamefont{K.~G.} \bibnamefont{Lee}},
  \bibinfo{author}{\bibfnamefont{X.~W.} \bibnamefont{Chen}},
  \bibinfo{author}{\bibfnamefont{H.}~\bibnamefont{Eghlidi}},
  \bibinfo{author}{\bibfnamefont{P.}~\bibnamefont{Kukura}},
  \bibinfo{author}{\bibfnamefont{R.}~\bibnamefont{Lettow}},
  \bibinfo{author}{\bibfnamefont{A.}~\bibnamefont{Renn}},
  \bibinfo{author}{\bibfnamefont{V.}~\bibnamefont{Sandoghdar}},
  \bibnamefont{and}
  \bibinfo{author}{\bibfnamefont{S.}~\bibnamefont{G{\"o}tzinger}},
  \bibinfo{journal}{Nature Photon.} \textbf{\bibinfo{volume}{5}},
  \bibinfo{pages}{166} (\bibinfo{year}{2011}).

\bibitem[{\citenamefont{Chang et~al.}(2006)\citenamefont{Chang, {S\o rensen},
  Hemmer, and Lukin}}]{Chang:06}
\bibinfo{author}{\bibfnamefont{D.~E.} \bibnamefont{Chang}},
  \bibinfo{author}{\bibfnamefont{A.~S.} \bibnamefont{{S\o rensen}}},
  \bibinfo{author}{\bibfnamefont{P.~R.} \bibnamefont{Hemmer}},
  \bibnamefont{and} \bibinfo{author}{\bibfnamefont{M.~D.} \bibnamefont{Lukin}},
  \bibinfo{journal}{Phys. Rev. Lett.} \textbf{\bibinfo{volume}{97}},
  \bibinfo{pages}{053002} (\bibinfo{year}{2006}).

\bibitem[{\citenamefont{Friedler et~al.}(2009)\citenamefont{Friedler, Sauvan,
  Hugonin, Lalanne, Claudon, and G\'erard}}]{Friedler:09}
\bibinfo{author}{\bibfnamefont{I.}~\bibnamefont{Friedler}},
  \bibinfo{author}{\bibfnamefont{C.}~\bibnamefont{Sauvan}},
  \bibinfo{author}{\bibfnamefont{J.~P.} \bibnamefont{Hugonin}},
  \bibinfo{author}{\bibfnamefont{P.}~\bibnamefont{Lalanne}},
  \bibinfo{author}{\bibfnamefont{J.}~\bibnamefont{Claudon}}, \bibnamefont{and}
  \bibinfo{author}{\bibfnamefont{J.~M.} \bibnamefont{G\'erard}},
  \bibinfo{journal}{Opt. Exp.} \textbf{\bibinfo{volume}{17}},
  \bibinfo{pages}{2095} (\bibinfo{year}{2009}).

\bibitem[{\citenamefont{Shen and Fan}(2005)}]{Shen:05}
\bibinfo{author}{\bibfnamefont{J.~T.} \bibnamefont{Shen}} \bibnamefont{and}
  \bibinfo{author}{\bibfnamefont{S.}~\bibnamefont{Fan}}, \bibinfo{journal}{Opt.
  Lett.} \textbf{\bibinfo{volume}{30}}, \bibinfo{pages}{2001}
  (\bibinfo{year}{2005}).

\bibitem[{\citenamefont{Hwang and Hinds}(2011)}]{hwang:11}
\bibinfo{author}{\bibfnamefont{J.}~\bibnamefont{Hwang}} \bibnamefont{and}
  \bibinfo{author}{\bibfnamefont{E.~A.} \bibnamefont{Hinds}},
  \bibinfo{journal}{New J. Phys.} \textbf{\bibinfo{volume}{13}},
  \bibinfo{pages}{085009} (\bibinfo{year}{2011}).

\bibitem[{\citenamefont{Vetsch et~al.}(2010)\citenamefont{Vetsch, Reitz,
  Sagu\'{e}, Schmidt, Dawkins, and Rauschenbeutel}}]{vetsch_optical_2010}
\bibinfo{author}{\bibfnamefont{E.}~\bibnamefont{Vetsch}},
  \bibinfo{author}{\bibfnamefont{D.}~\bibnamefont{Reitz}},
  \bibinfo{author}{\bibfnamefont{G.}~\bibnamefont{Sagu\'{e}}},
  \bibinfo{author}{\bibfnamefont{R.}~\bibnamefont{Schmidt}},
  \bibinfo{author}{\bibfnamefont{S.~T.} \bibnamefont{Dawkins}},
  \bibnamefont{and}
  \bibinfo{author}{\bibfnamefont{A.}~\bibnamefont{Rauschenbeutel}},
  \bibinfo{journal}{Phys. Rev. Lett.} \textbf{\bibinfo{volume}{104}},
  \bibinfo{pages}{203603} (\bibinfo{year}{2010}).

\bibitem[{\citenamefont{Yalla et~al.}(2012)\citenamefont{Yalla, Le~Kien,
  Morinaga, and Hakuta}}]{yalla_efficient_2012}
\bibinfo{author}{\bibfnamefont{R.}~\bibnamefont{Yalla}},
  \bibinfo{author}{\bibfnamefont{F.}~\bibnamefont{Le~Kien}},
  \bibinfo{author}{\bibfnamefont{M.}~\bibnamefont{Morinaga}}, \bibnamefont{and}
  \bibinfo{author}{\bibfnamefont{K.}~\bibnamefont{Hakuta}},
  \bibinfo{journal}{Phy. Rev. Lett.} \textbf{\bibinfo{volume}{109}},
  \bibinfo{pages}{063602} (\bibinfo{year}{2012}).

\bibitem[{\citenamefont{Akimov et~al.}(2007)\citenamefont{Akimov, Mukherjee,
  Yu, Chang, Zibrov, Hemmer, Park, and Lukin}}]{Akimov:07}
\bibinfo{author}{\bibfnamefont{A.~V.} \bibnamefont{Akimov}},
  \bibinfo{author}{\bibfnamefont{A.}~\bibnamefont{Mukherjee}},
  \bibinfo{author}{\bibfnamefont{C.~L.} \bibnamefont{Yu}},
  \bibinfo{author}{\bibfnamefont{D.~E.} \bibnamefont{Chang}},
  \bibinfo{author}{\bibfnamefont{A.~S.} \bibnamefont{Zibrov}},
  \bibinfo{author}{\bibfnamefont{P.~R.} \bibnamefont{Hemmer}},
  \bibinfo{author}{\bibfnamefont{H.}~\bibnamefont{Park}}, \bibnamefont{and}
  \bibinfo{author}{\bibfnamefont{M.~D.} \bibnamefont{Lukin}},
  \bibinfo{journal}{Nature} \textbf{\bibinfo{volume}{450}},
  \bibinfo{pages}{402} (\bibinfo{year}{2007}).

\bibitem[{\citenamefont{Lund-Hansen et~al.}(2008)\citenamefont{Lund-Hansen,
  Stobbe, Julsgaard, Thyrrestrup, S{\o}nner, Kamp, Forchel, and
  Lodahl}}]{lund-hansen_experimental_2008}
\bibinfo{author}{\bibfnamefont{T.}~\bibnamefont{Lund-Hansen}},
  \bibinfo{author}{\bibfnamefont{S.}~\bibnamefont{Stobbe}},
  \bibinfo{author}{\bibfnamefont{B.}~\bibnamefont{Julsgaard}},
  \bibinfo{author}{\bibfnamefont{H.}~\bibnamefont{Thyrrestrup}},
  \bibinfo{author}{\bibfnamefont{T.}~\bibnamefont{S{\o}nner}},
  \bibinfo{author}{\bibfnamefont{M.}~\bibnamefont{Kamp}},
  \bibinfo{author}{\bibfnamefont{A.}~\bibnamefont{Forchel}}, \bibnamefont{and}
  \bibinfo{author}{\bibfnamefont{P.}~\bibnamefont{Lodahl}},
  \bibinfo{journal}{Phys. Rev. Lett.} \textbf{\bibinfo{volume}{101}},
  \bibinfo{pages}{113903} (\bibinfo{year}{2008}).

\bibitem[{\citenamefont{Goban et~al.}(2014)\citenamefont{Goban, Hung, Yu, Hood,
  Muniz, Lee, Martin, McClung, Choi, Chang et~al.}}]{Goban:14}
\bibinfo{author}{\bibfnamefont{A.}~\bibnamefont{Goban}},
  \bibinfo{author}{\bibfnamefont{C.-L.} \bibnamefont{Hung}},
  \bibinfo{author}{\bibfnamefont{S.-P.} \bibnamefont{Yu}},
  \bibinfo{author}{\bibfnamefont{J.}~\bibnamefont{Hood}},
  \bibinfo{author}{\bibfnamefont{J.}~\bibnamefont{Muniz}},
  \bibinfo{author}{\bibfnamefont{J.}~\bibnamefont{Lee}},
  \bibinfo{author}{\bibfnamefont{M.}~\bibnamefont{Martin}},
  \bibinfo{author}{\bibfnamefont{A.}~\bibnamefont{McClung}},
  \bibinfo{author}{\bibfnamefont{K.}~\bibnamefont{Choi}},
  \bibinfo{author}{\bibfnamefont{D.}~\bibnamefont{Chang}},
  \bibnamefont{et~al.}, \bibinfo{journal}{Nat. Commun.}
  \textbf{\bibinfo{volume}{5}}, \bibinfo{pages}{3808} (\bibinfo{year}{2014}).

\bibitem[{\citenamefont{Astafiev et~al.}(2010)\citenamefont{Astafiev, Zagoskin,
  Abdumalikov, Pashkin, Yamamoto, Inomata, Nakamura, and Tsai}}]{Astafiev:10}
\bibinfo{author}{\bibfnamefont{O.}~\bibnamefont{Astafiev}},
  \bibinfo{author}{\bibfnamefont{A.~M.} \bibnamefont{Zagoskin}},
  \bibinfo{author}{\bibfnamefont{A.~A.} \bibnamefont{Abdumalikov}},
  \bibinfo{author}{\bibfnamefont{Y.~A.} \bibnamefont{Pashkin}},
  \bibinfo{author}{\bibfnamefont{T.}~\bibnamefont{Yamamoto}},
  \bibinfo{author}{\bibfnamefont{K.}~\bibnamefont{Inomata}},
  \bibinfo{author}{\bibfnamefont{Y.}~\bibnamefont{Nakamura}}, \bibnamefont{and}
  \bibinfo{author}{\bibfnamefont{J.~S.} \bibnamefont{Tsai}},
  \bibinfo{journal}{Science} \textbf{\bibinfo{volume}{327}},
  \bibinfo{pages}{840} (\bibinfo{year}{2010}).

\bibitem[{\citenamefont{Moerner and Kador}(1989)}]{Moerner:89}
\bibinfo{author}{\bibfnamefont{W.~E.} \bibnamefont{Moerner}} \bibnamefont{and}
  \bibinfo{author}{\bibfnamefont{L.}~\bibnamefont{Kador}},
  \bibinfo{journal}{Phys. Rev. Lett.} \textbf{\bibinfo{volume}{62}},
  \bibinfo{pages}{2535} (\bibinfo{year}{1989}).

\bibitem[{\citenamefont{Jelezko et~al.}(1996)\citenamefont{Jelezko, Tamarat,
  Lounis, and Orrit}}]{Jelezko96}
\bibinfo{author}{\bibfnamefont{F.}~\bibnamefont{Jelezko}},
  \bibinfo{author}{\bibfnamefont{P.}~\bibnamefont{Tamarat}},
  \bibinfo{author}{\bibfnamefont{B.}~\bibnamefont{Lounis}}, \bibnamefont{and}
  \bibinfo{author}{\bibfnamefont{M.}~\bibnamefont{Orrit}}, \bibinfo{journal}{J.
  Phys. Chem.} \textbf{\bibinfo{volume}{100}}, \bibinfo{pages}{13892}
  (\bibinfo{year}{1996}).

\bibitem[{\citenamefont{Haakh et~al.}(2014)\citenamefont{Haakh, Faez, and
  Sandoghdar}}]{Haakh:14}
\bibinfo{author}{\bibfnamefont{H.}~\bibnamefont{Haakh}},
  \bibinfo{author}{\bibfnamefont{S.}~\bibnamefont{Faez}}, \bibnamefont{and}
  \bibinfo{author}{\bibfnamefont{V.}~\bibnamefont{Sandoghdar}},
  \bibinfo{journal}{in preparation}  (\bibinfo{year}{2014}).

\bibitem[{EPA()}]{EPAPS-Tuerschmann}
\bibinfo{note}{See online supplementar material (SM) for an intuitive
  derivation of the relationship between the extinction dip and $\beta$ as well
  as the distribution of the resonance linewidths in the nanocapillary.}

\bibitem[{\citenamefont{Basche et~al.}(1999)\citenamefont{Basche, Moerner,
  Orrit, and Wild}}]{SMbook}
\bibinfo{author}{\bibfnamefont{T.}~\bibnamefont{Basche}},
  \bibinfo{author}{\bibfnamefont{W.~E.} \bibnamefont{Moerner}},
  \bibinfo{author}{\bibfnamefont{M.}~\bibnamefont{Orrit}}, \bibnamefont{and}
  \bibinfo{author}{\bibfnamefont{U.}~\bibnamefont{Wild}},
  \emph{\bibinfo{title}{Single Molecule Spectroscopy}}
  (\bibinfo{publisher}{John Wiley and Sons}, \bibinfo{year}{1999}).

\bibitem[{\citenamefont{Walther}(2005)}]{Walther:05}
\bibinfo{author}{\bibfnamefont{H.}~\bibnamefont{Walther}}, in
  \emph{\bibinfo{booktitle}{Advances in Atomic, Molecular and Optical Physics,
  Vol. 51}} (\bibinfo{publisher}{Elsevier}, \bibinfo{year}{2005}), pp.
  \bibinfo{pages}{239--272}.

\bibitem[{\citenamefont{S.Weisenburger
  et~al.}(2014)\citenamefont{S.Weisenburger, Jing, H\"anni, Reymond, Schuler,
  Renn, and Sandoghdar}}]{Weisenburger:14}
\bibinfo{author}{\bibnamefont{S.Weisenburger}},
  \bibinfo{author}{\bibfnamefont{B.}~\bibnamefont{Jing}},
  \bibinfo{author}{\bibfnamefont{D.}~\bibnamefont{H\"anni}},
  \bibinfo{author}{\bibfnamefont{L.}~\bibnamefont{Reymond}},
  \bibinfo{author}{\bibfnamefont{B.}~\bibnamefont{Schuler}},
  \bibinfo{author}{\bibfnamefont{A.}~\bibnamefont{Renn}}, \bibnamefont{and}
  \bibinfo{author}{\bibfnamefont{V.}~\bibnamefont{Sandoghdar}},
  \bibinfo{journal}{ChemPhysChem} \textbf{\bibinfo{volume}{15}},
  \bibinfo{pages}{763} (\bibinfo{year}{2014}).

\bibitem[{\citenamefont{Gonzalez-Tudela
  et~al.}(2011)\citenamefont{Gonzalez-Tudela, Martin-Cano, Moreno,
  Martin-Moreno, Tejedor, and Garcia-Vidal}}]{Gonzalez-Tudela:11}
\bibinfo{author}{\bibfnamefont{A.}~\bibnamefont{Gonzalez-Tudela}},
  \bibinfo{author}{\bibfnamefont{D.}~\bibnamefont{Martin-Cano}},
  \bibinfo{author}{\bibfnamefont{E.}~\bibnamefont{Moreno}},
  \bibinfo{author}{\bibfnamefont{L.}~\bibnamefont{Martin-Moreno}},
  \bibinfo{author}{\bibfnamefont{C.}~\bibnamefont{Tejedor}}, \bibnamefont{and}
  \bibinfo{author}{\bibfnamefont{F.~J.} \bibnamefont{Garcia-Vidal}},
  \bibinfo{journal}{Phys. Rev. Lett.} \textbf{\bibinfo{volume}{106}},
  \bibinfo{pages}{020501} (\bibinfo{year}{2011}).

\bibitem[{\citenamefont{Hettich et~al.}(2002)\citenamefont{Hettich, Schmitt,
  Zitzmann, K\"{u}hn, Gerhardt, and Sandoghdar}}]{Hettich:02}
\bibinfo{author}{\bibfnamefont{C.}~\bibnamefont{Hettich}},
  \bibinfo{author}{\bibfnamefont{C.}~\bibnamefont{Schmitt}},
  \bibinfo{author}{\bibfnamefont{J.}~\bibnamefont{Zitzmann}},
  \bibinfo{author}{\bibfnamefont{S.}~\bibnamefont{K\"{u}hn}},
  \bibinfo{author}{\bibfnamefont{I.}~\bibnamefont{Gerhardt}}, \bibnamefont{and}
  \bibinfo{author}{\bibfnamefont{V.}~\bibnamefont{Sandoghdar}},
  \bibinfo{journal}{Science} \textbf{\bibinfo{volume}{298}},
  \bibinfo{pages}{385} (\bibinfo{year}{2002}).

\bibitem[{\citenamefont{Rezus et~al.}(2012)\citenamefont{Rezus, Walt, Lettow,
  Renn, Zumofen, G\"otzinger, and Sandoghdar}}]{Rezus:12}
\bibinfo{author}{\bibfnamefont{Y.~L.~A.} \bibnamefont{Rezus}},
  \bibinfo{author}{\bibfnamefont{S.~G.} \bibnamefont{Walt}},
  \bibinfo{author}{\bibfnamefont{R.}~\bibnamefont{Lettow}},
  \bibinfo{author}{\bibfnamefont{A.}~\bibnamefont{Renn}},
  \bibinfo{author}{\bibfnamefont{G.}~\bibnamefont{Zumofen}},
  \bibinfo{author}{\bibfnamefont{S.}~\bibnamefont{G\"otzinger}},
  \bibnamefont{and}
  \bibinfo{author}{\bibfnamefont{V.}~\bibnamefont{Sandoghdar}},
  \bibinfo{journal}{Phys. Rev. Lett.} \textbf{\bibinfo{volume}{108}},
  \bibinfo{pages}{093601} (\bibinfo{year}{2012}).

\bibitem[{\citenamefont{van Loo et~al.}(2013)\citenamefont{van Loo, Fedorov,
  Lalumiere, Sanders, Blais, and Wallraff}}]{vanLoo:13}
\bibinfo{author}{\bibfnamefont{A.~F.} \bibnamefont{van Loo}},
  \bibinfo{author}{\bibfnamefont{A.}~\bibnamefont{Fedorov}},
  \bibinfo{author}{\bibfnamefont{K.}~\bibnamefont{Lalumiere}},
  \bibinfo{author}{\bibfnamefont{B.~C.} \bibnamefont{Sanders}},
  \bibinfo{author}{\bibfnamefont{A.}~\bibnamefont{Blais}}, \bibnamefont{and}
  \bibinfo{author}{\bibfnamefont{A.}~\bibnamefont{Wallraff}},
  \bibinfo{journal}{Science} \textbf{\bibinfo{volume}{342}},
  \bibinfo{pages}{1494} (\bibinfo{year}{2013}).

\bibitem[{\citenamefont{Gerhardt et~al.}(2009)\citenamefont{Gerhardt, Wrigge,
  Zumofen, Hwang, Renn, and Sandoghdar}}]{Gerhardt:09}
\bibinfo{author}{\bibfnamefont{I.}~\bibnamefont{Gerhardt}},
  \bibinfo{author}{\bibfnamefont{G.}~\bibnamefont{Wrigge}},
  \bibinfo{author}{\bibfnamefont{G.}~\bibnamefont{Zumofen}},
  \bibinfo{author}{\bibfnamefont{J.}~\bibnamefont{Hwang}},
  \bibinfo{author}{\bibfnamefont{A.}~\bibnamefont{Renn}}, \bibnamefont{and}
  \bibinfo{author}{\bibfnamefont{V.}~\bibnamefont{Sandoghdar}},
  \bibinfo{journal}{Phys. Rev. A.} \textbf{\bibinfo{volume}{79}},
  \bibinfo{pages}{011402(R)} (\bibinfo{year}{2009}).

\bibitem[{\citenamefont{Hwang et~al.}(2009)\citenamefont{Hwang, Pototschnig,
  Lettow, Zumofen, Renn, G\"otzinger, and Sandoghdar}}]{Hwang:09}
\bibinfo{author}{\bibfnamefont{J.}~\bibnamefont{Hwang}},
  \bibinfo{author}{\bibfnamefont{M.}~\bibnamefont{Pototschnig}},
  \bibinfo{author}{\bibfnamefont{R.}~\bibnamefont{Lettow}},
  \bibinfo{author}{\bibfnamefont{G.}~\bibnamefont{Zumofen}},
  \bibinfo{author}{\bibfnamefont{A.}~\bibnamefont{Renn}},
  \bibinfo{author}{\bibfnamefont{S.}~\bibnamefont{G\"otzinger}},
  \bibnamefont{and}
  \bibinfo{author}{\bibfnamefont{V.}~\bibnamefont{Sandoghdar}},
  \bibinfo{journal}{Nature} \textbf{\bibinfo{volume}{460}}, \bibinfo{pages}{76}
  (\bibinfo{year}{2009}).

\bibitem[{\citenamefont{Shen and Fan}(2007)}]{Shen:07}
\bibinfo{author}{\bibfnamefont{J.-T.} \bibnamefont{Shen}} \bibnamefont{and}
  \bibinfo{author}{\bibfnamefont{S.}~\bibnamefont{Fan}},
  \bibinfo{journal}{Phys. Rev. Lett.} \textbf{\bibinfo{volume}{98}},
  \bibinfo{pages}{153003} (\bibinfo{year}{2007}).

\bibitem[{\citenamefont{Greentree et~al.}(2006)\citenamefont{Greentree, Tahan,
  Cole, and Hollenberg}}]{Greentree:06}
\bibinfo{author}{\bibfnamefont{A.~D.} \bibnamefont{Greentree}},
  \bibinfo{author}{\bibfnamefont{C.}~\bibnamefont{Tahan}},
  \bibinfo{author}{\bibfnamefont{J.~H.} \bibnamefont{Cole}}, \bibnamefont{and}
  \bibinfo{author}{\bibfnamefont{L.~C.~L.} \bibnamefont{Hollenberg}},
  \bibinfo{journal}{Nature Phys.} \textbf{\bibinfo{volume}{2}},
  \bibinfo{pages}{856} (\bibinfo{year}{2006}).

\bibitem[{\citenamefont{Quan et~al.}(2009)\citenamefont{Quan, Bulu, and
  Lon\v{c}ar}}]{Quan:09}
\bibinfo{author}{\bibfnamefont{Q.}~\bibnamefont{Quan}},
  \bibinfo{author}{\bibfnamefont{I.}~\bibnamefont{Bulu}}, \bibnamefont{and}
  \bibinfo{author}{\bibfnamefont{M.}~\bibnamefont{Lon\v{c}ar}},
  \bibinfo{journal}{Phys. Rev. A} \textbf{\bibinfo{volume}{80}},
  \bibinfo{pages}{011810(R)} (\bibinfo{year}{2009}).

\end{thebibliography}
\end{document}